\documentstyle[A4,epsf,12pt]{article}

\input{psfig}

\def\TM{T_{12345}}
\def\DTM{\Delta T_{12345}}
\def\Im{{\rm Im}}
\def\rmF{{\rm F}}
\def\ieps{i \epsilon}
\def\nn{ \nonumber \\ &&}

\begin{document}

\title{
\begin{flushright} {\large UWITP  -- 06  /94\\
                           hep-ph/9501201 \\[3ex] } \end{flushright}
Simple one-dimensional integral representations for two-loop self-energies:
the master diagram}
\author{
 \renewcommand{\thefootnote}{\fnsymbol{footnote}}
 S. Bauberger\footnotemark[1] $\,$,
 M. B\"ohm\footnotemark[2] \\[2ex]
  {\normalsize \it
      Institut f\"ur Theoretische Physik, Universit\"at W\"urzburg,}\\
  {\normalsize \it
            Am Hubland, D-97074 W\"urzburg, Germany} }
\date{December 1994}
\maketitle

\renewcommand{\thefootnote}{\fnsymbol{footnote}}
\footnotetext[1]
  {E-mail address: bauberger@physik.uni-wuerzburg.de}
\footnotetext[2]
  {E-mail address: boehm@physik.uni-wuerzburg.de}

\begin{abstract}
The scalar two-loop self-energy master diagram is studied in the case of
arbitrary masses. Analytical results in terms of Lauricella- and
Appell-functions are presented for the
imaginary part. By using the dispersion relation
a one-dimensional integral representation is derived.
This representation uses
only elementary functions and is thus well suited for a numerical
calculation of the master diagram.
\end{abstract}

\section{Introduction}

The recent years have led to a considerable progress concerning the precision
of measurements of some parameters of the standard
model. Outstanding examples are the determination of the the $Z$-mass and
-width \cite{MZ}. Also a precise determination of the top-mass can be expected
in the next years.
As a consequence, very precise theoretical evaluations of
the theory are more and more required, including
calculations of two-loop corrections.

This paper focuses on the evaluation of the so-called scalar master diagram,
which belongs to the class of two-loop self-energy diagrams.
Analytical solutions, involving generalized hypergeometric functions,
have been presented for all other scalar two-loop self-energy diagrams
in the general mass case \cite{Buza}-\cite{BBBB2}. But no comparable results
have so far been found for the master diagram, whereas
for special mass configurations or in asymptotic parameter regions there exist
many solutions \cite{Broadhurst}-\cite{asymptotic}.
In this paper we are concerned with the general mass case.

After an introduction of our notations in sect.~\ref{notation},
we investigate in sect.~\ref{MasterImag}
the imaginary part of the master diagram. Analytical results are presented
for the two-particle cut contributions and three-particle cut contributions.

Concerning the full calculation of the master integral, it must be noted
\cite{Scharf} that for two-loop integrals which contain a massive
three-particle cut, the results in general cannot be expressed
in terms of generalized polylogarithms with algebraic functions
of the parameters as arguments.
Thus, a promising approach to two-loop integrals uses
integral representations which involve only elementary functions in
the integrand and can therefore easily be evaluated numerically.
For the master diagram a
two-dimensional integral representation \cite{Kreimer} has become
famous, which involves only simple logarithms and square roots.
This approach has been generalized to other two-loop self-energy integrals
in ref.~\cite{Berends}.

In sect.~\ref{one-dimensional-representations}
we present a one-dimensional integral representation which
only involves elementary functions. The standard algorithms for the
numerical evaluation of one-di\-men\-sio\-nal integrals show
very good convergence properties. Therefore the representation presented
in this paper
allows a very fast numerical calculation of the master diagram.

In sect.~\ref{Numerics} we present some numerical comparisons
of our results with those of \cite{Kreimer}, demonstrating the agreement
of the results obtained with both methods.

\section{Notation and definitions}
\label{notation}

\begin{figure}[htb]
\unitlength1cm
\epsfysize=4cm
\centerline{\epsffile{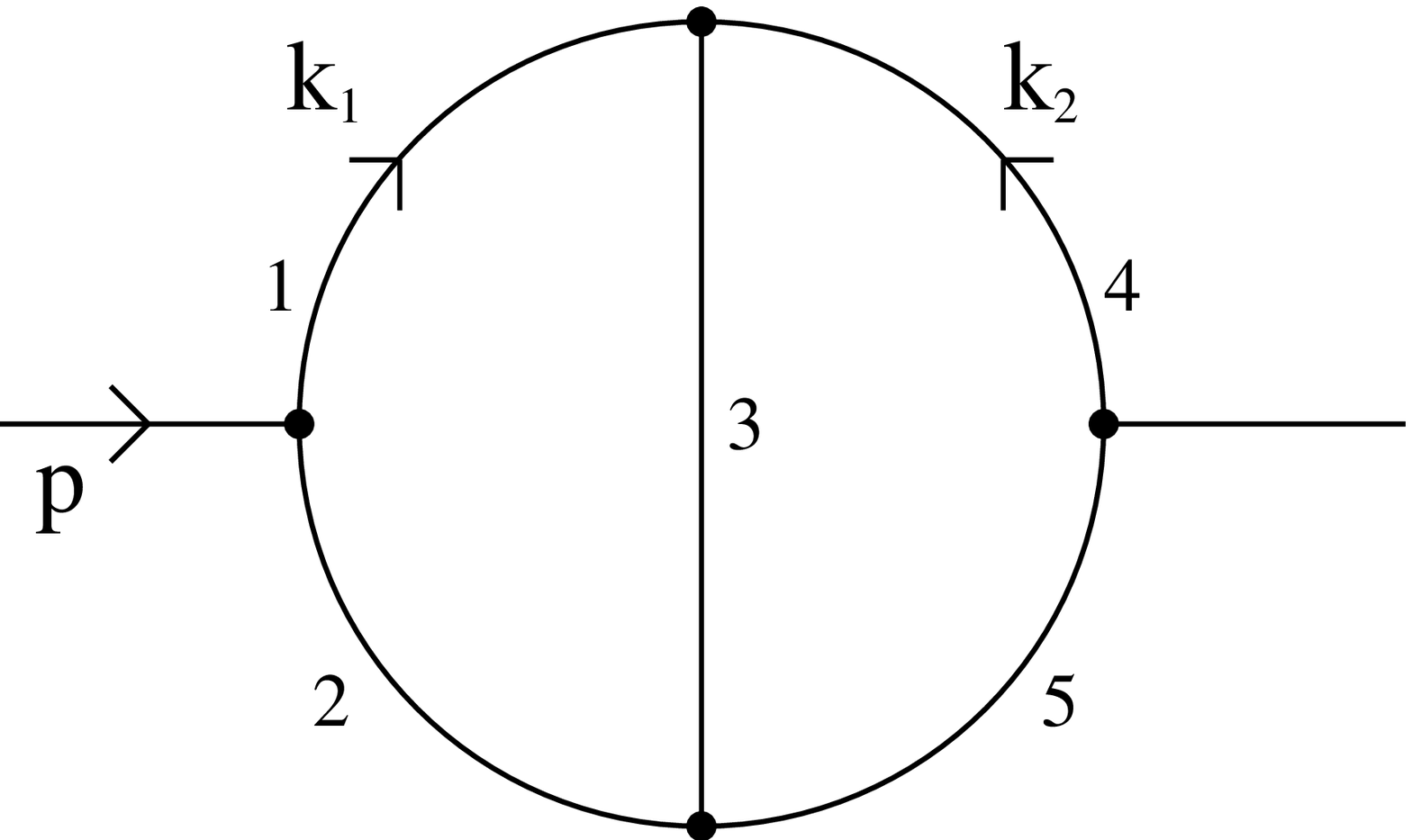}}
\caption{The master diagram $\TM$.} \label{Master}
\end{figure}

The master integral (fig.~\ref{Master}) is in the general mass case
determined by six independent parameters, i.e.~five
internal masses and the external squared momentum $p^2$. The name 'master
diagram' has been introduced by Broadhurst and is due to the fact
that a cancellation of propagators in this diagram
generates all other fundamental two-loop self-energy diagrams.
The integral $\TM(p^2;m_1^2,m_2^2,m_3^2,m_4^2,m_5^2)$ is finite
in four dimensions.
We use the convention
\begin{eqnarray}
    \TM(p^2;m_1^2,m_2^2,m_3^2,m_4^2,m_5^2) &\!\!=&\!\!
    -\frac{1}{\pi^4} \int d^4 k_1 \, d^4 k_2 \,
         \frac{1}{(k_1^2-m_1^2)((k_1-p)^2-m_2^2)} \nn
    \;  \times
   \frac{1}{((k_1+k_2)^2-m_3^2)(k_2^2-m_4^2)((k_2+p)^2-m_5^2)} \, ,
\end{eqnarray}
where all masses have an infinitely small negative imaginary part.

\begin{figure}[htb]
\unitlength1cm
\epsfxsize=\textwidth
\centerline{\epsffile{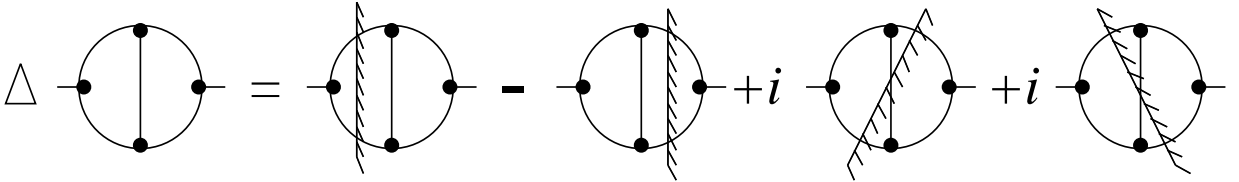}}
\caption{The contributions to the discontinuity of the master diagram.}
\label{Master-contrib}
\end{figure}

The integral $\TM$ has branch cuts along the positive real axis.
The discontinuity across these cuts is related to the imaginary part,
\begin{equation}
  \DTM(p^2;m_i^2) = 2i \, \Im \left( \TM(p^2;m_i^2) \right) \,,  \nonumber
\end{equation}
where $m_i^2$ denotes the mass parameters.
The integral can be calculated from the discontinuity
with the dispersion relation
\begin{equation}
  \TM(p^2;m_i^2)= \frac{1}{2 \pi i} \int\limits_{s_0}^{\infty} ds \,
                        \frac{\DTM(s;m_i^2)}{s-p^2-\ieps}  \, ,
   \label{dispersion}
\end{equation}
where $s_0$ denotes the lowest branch point of the function.
According to the Cutkosky rules \cite{Cutkosky,Itzykson},
the discontinuity of the master diagram can be decomposed into four parts,
each corresponding to a cut diagram (fig.~\ref{Master-contrib}),
\begin{equation}
  \DTM = \DTM^{(2a)} + \DTM^{(2b)} + \DTM^{(3a)} + \DTM^{(3b)} \, ,
     \nonumber
\end{equation}
where $\DTM^{(2a)}$, $\DTM^{(2b)}$ and  $\DTM^{(3a)}$, $\DTM^{(3b)}$
denote the two- and three-particle discontinuities,
cf.~fig.~\ref{Master-contrib}.

Each cut gives a contribution to the dispersion integral,
which will be denoted
accordingly,
\begin{equation}
  \TM^{(2a)}(p^2;m_i^2)=\frac{1}{2 \pi i} \int\limits_{s_0^{(2a)}}^\infty
    ds \,  \frac{\DTM^{(2a)}(s;m_i^2)}{s-p^2-\ieps} \,  .
    \label{T-contrib-2a}
\end{equation}

For each contribution, the lower limit of the integration is
determined by the threshold which is associated with the cut,
i.e.~$s_0^{(2a)}=(m_1+m_2)^2$, $s_0^{(2b)}=(m_4+m_5)^2$,
$s_0^{(3a)}=(m_2+m_3+m_4)^2$, $s_0^{(3b)}=(m_1+m_3+m_5)^2$.

\section{Calculation of the imaginary part}
\label{MasterImag}

\subsection{Explicit results for the two-particle discontinuities}

\begin{figure}[htb]
\unitlength1cm
\epsfysize=3.0cm
\centerline{\epsffile{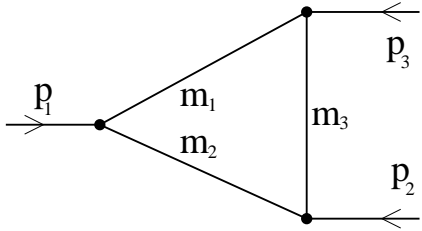}}
\caption{The triangle diagram.}
\label{C0fig}
\end{figure}

The two-particle cut contributions to the discontinuity are given according to
the Cutkosky rules as
\begin{eqnarray}
  \DTM^{(2a)} &=& -
 \frac{1}{\pi^4} \int d^4 k_1 \, d^4 k_2 \,
              2 \pi \Theta (k_1^0) \delta (k_1^2-m_1^2)
              2 \pi \Theta (-k_1^0+p^0) \delta ((k_1-p)^2-m_2^2)
        \nonumber \\  && \;\;\;\;\;\;\;
        \times \frac{1}{((k_1+k_2)^2-m_3^2-\ieps)
       (k_2^2-m_4^2-\ieps)((k_2+p)^2-m_5^2-\ieps)} \, ,
   \label{DT2a1} \\
  \DTM^{(2b)} &=& \frac{1}{\pi^4} \int d^4 k_1 \, d^4 k_2 \,
               2 \pi \Theta (-k_2^0) \delta (k_2^2-m_4^2)
               2 \pi \Theta (k_2^0+p^0) \delta ((k_2+p)^2-m_5^2)
        \nonumber \\   && \;\;\;\;\;\;
        \times \frac{1}{(k_1^2-m_1^2+\ieps)
         ((k_1-p)^2-m_2^2+\ieps)((k_1+k_2)^2-m_3^2+\ieps)} \, .
       \label{DT2b1}
\end{eqnarray}
In the evaluation of the expressions (\ref{DT2a1}) and (\ref{DT2b1}),
the discontinuity of the one-loop self-energy in four dimensions,
\begin{eqnarray}
  \Delta B_0(p^2;m_1^2,m_2^2) &=& \frac{i}{\pi^2} \int d^4 k \,
      2 \pi \Theta (k^0) \delta (k^2-m_1^2)
      2 \pi \Theta (p^0-k^0) \delta \left((p-k)^2-m_2^2 \right)
    \nonumber \\
    &=& 2 \pi i \, \frac{\sqrt{\lambda(p^2,m_1^2,m_2^2)}}{p^2} \,
         \Theta(p^2-(m_1+m_2)^2) \, , \label{DeltaB0}
\end{eqnarray}
occurs, where the K\"{a}ll\'{e}n function $\lambda$ is defined by
$\lambda(x,y,z)=(x-y-z)^2- 4 y z$.
The expressions (\ref{DT2a1}) and (\ref{DT2b1}) yield
\begin{eqnarray}
   \DTM^{(2a)} &=& \frac{i}{\pi^2} \int d^4 k_1 \,
              2 \pi \Theta (k_1^0) \delta (k_1^2-m_1^2)
              2 \pi \Theta (p^0-k_1^0) \delta ((k_1-p)^2-m_2^2)
        \nonumber \\
  && \;\;\;\;\;\;\;\; \times
   \left( C_0(p^2, k_1^2 , (k_1-p)^2 ;
         m_5^2,m_4^2, m_3^2) \right)^*  \nonumber \\
  &=& \Delta B_0(p^2;m_1^2,m_2^2)
      \left( C_0 (p^2,m_1^2,m_2^2; m_5^2,m_4^2, m_3^2) \right) ^* \, ,
  \label{DT2a}
\end{eqnarray}
and equivalently,
\begin{equation}
   \DTM^{(2b)} = \Delta B_0(p^2;m_4^2,m_5^2)
      C_0 (p^2,m_4^2,m_5^2; m_2^2,m_1^2, m_3^2) \, ,
  \label{DT2b}
\end{equation}
where $C_0$ denotes the integral associated with the one-loop triangle diagram
(fig.~\ref{C0fig}),
whose analytical solution involves dilogarithms \cite{tHooft2}.
The two-particle discontinuities $\DTM^{(2a)}$ and $\DTM^{(2b)}$ are hence
given by a product of the one-loop self-energy discontinuity $\Delta B_0$ and
the one-loop triangle integral $C_0$ or its complex conjugate $C_0^*$.

\subsection{Integral representation for the three-particle discontinuity}

For the three-particle discontinuities $\DTM^{(3a)}$
the Cutkosky rules yield
\begin{eqnarray}
  &&\!\!\!\!\!\!\!\! \DTM^{(3a)}(p^2;m_1^2,m_2^2,m_3^2,m_4^2,m_5^2) \nonumber
  \nn  =
  - \frac{i}{\pi^4} \int d^4 k_1 \, d^4 k_2 \,
         2 \pi \Theta (-k_1^0+p^0) \delta ((k_1-p)^2-m_2^2)
         2 \pi \Theta (k_1^0+k_2^0) \delta ((k_1+k_2)^2-m_3^2)
    \nonumber \\ && \quad \quad \quad \quad
         \times 2 \pi \Theta (-k_2^0) \delta (k_2^2-m_4^2)
        \frac{1}{(k_1^2-m_1^2+\ieps)((k_2+p)^2-m_5^2-\ieps)} \, .
      \label{DT3a1}
\end{eqnarray}
The contribution  $\DTM^{(3b)}$ requires no separate evaluation. It is
given by
\begin{equation}
   \DTM^{(3b)}(p^2;m_1^2,m_2^2,m_3^2,m_4^2,m_5^2) =
   \DTM^{(3a)}(p^2;m_2^2,m_1^2,m_3^2,m_5^2,m_4^2) \, . \nonumber
\end{equation}
With the dispersion representation of the propagator,
\begin{equation}
  \frac{1}{k_1^2-m_1^2+i\epsilon} = \int\limits_{-\infty}^{\infty} \, dt \,
                          \frac{\delta(k_1^2-t)}{t-m_1^2+i\epsilon} \, ,
\end{equation}
the two momentum integrations in (\ref{DT3a1}) can be performed
separately and yield
\begin{equation}
  \DTM^{(3a)}(p^2;m_i^2)
    = - \frac{1}{2 \pi i}
    \int\limits_{(m_3+m_4)^2}^{(\sqrt{p^2}-m_2)^2} \!\! dt \,
    \frac{\Delta B_0(p^2;t,m_2^2) \, \Delta C_0(t,m_2^2,p^2;m_4^2,m_3^2,m_5^2)}
         {t-m_1^2+i\epsilon} \, . \label{DT3}
\end{equation}
The lower limit of the integration follows from the threshold of the
$\Delta B_0$-function (\ref{DeltaB0}), while the upper limit guarantees
that $p^2>(m_2+m_3+m_4)^2$, i.e.~that $p^2$ is larger then the
three-particle threshold.
In expression (\ref{DT3}), the function $\Delta C_0$ denotes
the discontinuity of the one-loop integral $C_0$
(fig.~\ref{C0fig}) with respect to the parameter $p_1^2$.

This discontinuity is the main ingredient in the dispersion representation
of $C_0$, which will be used in sect.~\ref{one-dimensional-representations}.
The function $C_0$ can be calculated as
\begin{eqnarray}
  C_0(p_1^2,p_2^2,p_3^2;m_1^2,m_2^2,m_3^2)
        &=&\frac{1}{2\pi i} \int\limits_{s_0} ^\infty
      \frac{ \Delta C_0(s,p_2^2,p_3^2;m_1^2,m_2^2,m_3^2)}
          {s-p_1^2-i\epsilon} ds  \nonumber \\
  &&
       + C_{0an}(p_1^2,p_2^2,p_3^2;m_1^2,m_2^2,m_3^2)
            \, , \label{dispC0}
\end{eqnarray}
where the lower integration limit $s_0=(m_1+m_2)^2$ is the threshold belonging
to $p_1^2$, $\Delta C_0$ is the discontinuity associated with the normal
threshold and $C_{0an}$ is a contribution of an anomalous threshold.
The discontinuity $\Delta C_0$ is in general given by
\begin{eqnarray}
  && \Delta C_0(s,p_2^2,p_3^2;m_1^2,m_2^2,m_3^2)  \nonumber \\
  &&  \;\;\; = \frac{i}{\pi^2} \int d^4 k \,
       2 \pi \Theta (k^0) \delta (k^2 -m_1^2) \,
       2 \pi \Theta (-k^0-p_2^0-p_3^0)
             \delta ((k+p_2+p_3)^2 -m_2^2) \nonumber \\
  && \quad \quad \quad \times \frac{1}{(k+p_3)^2-m_3^2-\ieps} \nonumber \\
  && \;\;\;  = - 2 \pi i \, \frac{\log(a+b-\ieps)-\log(a-b-\ieps)}
                {\sqrt{(\Sigma p)^2-s+\ieps} \,
                 \sqrt{(\Delta p)^2-s+\ieps}}
                    \label{DeltaC0} \,
                 \Theta(s-(m_1+m_2)^2) \\
  && \;\;\; =  - 2 \pi i \, \frac{2 \tanh^{-1} (b/a)}
                {\sqrt{(\Sigma p)^2-s+\ieps} \,
                 \sqrt{(\Delta p)^2-s+\ieps}}
                 \,\Theta(s-(m_1+m_2)^2) \,,
\end{eqnarray}
where
\begin{eqnarray}
  && a\,(s,p_2^2,p_3^2;m_1^2,m_2^2,m_3^2) \nonumber \\
  && \;\;\; =s\,(s + 2 m_3^2 - p_2^2 -p_3^2 - m_1^2 - m_2^2)
        + (p_3^2-p_2^2) (m_1^2-m_2^2) \;,
     \label{a} \\
  && b\,(s,p_2^2,p_3^2;m_1^2,m_2^2) =
                \sqrt{(\Sigma p)^2-s+\ieps} \,
                \sqrt{(\Delta p)^2-s+\ieps} \,
                \sqrt{\lambda(s,m_1^2,m_2^2)} \, , \label{b} \\
  && \Sigma p = |p_2| + |p_3| \, ,
     \quad \Delta p = |p_2| - |p_3|  \,, \nonumber
\end{eqnarray}
and it is assumed that $p_2^2>0$ and $p_3^2>0$.

In order to produce a reliable numerical program for the master diagram,
a careful discussion of the analytical properties of $C_0$ is required.
The Landau equations show that a leading singularity and thus an anomalous
threshold occurs in $C_0$ if
\begin{equation}
  m_3^2 \,<\, m_{3thr}^2= \frac{m_1 p_2^2 + m_2 p_3^2}{m_1+m_2} - m_1 m_2 \, .
     \label{m3thr}
\end{equation}
The branch point of this anomalous threshold is given by
\begin{eqnarray}
  && \!\!\!\!\!\!\!\!\!\!\!\!
   s_1= \frac{p_2^2+p_3^2+m_1^2+m_2^2-m_3^2}{2}
          - \frac{(p_2^2-m_2^2)(p_3^2-m_1^2)
                   +l\,(p_2^2,m_2^2,m_3^2)\,l\,(p_3^2,m_1^2,m_3^2)}
                 {2\, m_3^2}  \, , \label{s1}  \\
  && \!\!\!\!\!\!\!\!\!\!\!\! \mbox{with} \quad l\,(p_i^2,m_j^2,m_k^2)
          := \sqrt{p_i^2-(m_j-m_k)^2+i\epsilon} \,
                  \sqrt{p_i^2-(m_j+m_k)^2+i\epsilon} \, . \nonumber
\end{eqnarray}
One can understand the effect of the anomalous threshold by an inspection
of the behavior of the branch point $s_1$ as a function of $m_3^2$,
taking into account that this parameter has an infinitely small imaginary part.
In terms of the dispersion representation (\ref{dispC0}),
in which $C_{0an}$ vanishes for $m_3^2>m_{3thr}^2$, the threshold $s_1$ is a
singularity of the logarithm in $\Delta C_0$ as function of $s$.
At $m_3^2=m_{3thr}^2$ this
singularity is equal to $s_0$. For $m_3^2<m_{3thr}^2$ the singularity $s_1$
crosses in some cases the integration contour, which consequently has to be
deformed.
The additional contribution $C_{0an}$ is therefore
given by an integration over the discontinuity of the logarithm
\cite{anomalous},
\begin{equation}
   C_{0an} = \frac{1}{2\pi i} \int\limits_{s_0}^{s_1}
               \frac{(2\pi i)^2}
                    {\sqrt{\lambda(s,p_2^2,p_3^2)}}  \,
               \frac{ds}{s-p_1^2} \, , \label{C0an}
\end{equation}
which evaluates to square roots and logarithms,
\begin{eqnarray}
    && \!\!\!\!\!\!
 \int \frac{ds}{\sqrt{\lambda(s,p_2^2,p_3^2)} (s-p_1^2)}
   = \frac{1}{\sqrt{\lambda(p_1^2,p_2^2,p_3^2)}} \Big\{
        \log(p_1^2-s+i\epsilon) \nn
  \quad \quad \quad \quad   \quad
      - \log\big(-s(p_2^2+p_3^2-p_1^2)+(p_2^2-p_3^2)^2-
                       p_1^2(p_2^2+p_3^2) \nn
  \quad \quad \quad \quad  \quad \quad \quad \quad
               +  \sqrt{(\Sigma p)^2-s+i\epsilon}
                  \sqrt{(\Delta p)^2-s+i\epsilon}
                  \sqrt{\lambda(p_1^2,p_2^2,p_3^2)}
               + i \epsilon \big)
                          \Big\}
              \label{C0an-2}   \, .
\end{eqnarray}
{}From equation (\ref{C0an}) it can be observed that
for $s_1$ real and $s_1>s_0$ no separate treatment of
$C_{0an}$ is required, but the discontinuity of the logarithm
can be included as additional term to the logarithm in the discontinuity
$\Delta C_0$ (\ref{DeltaC0}).

Thus, one can distinguish the following six cases:
\begin{itemize}
\item {\em First case:} If $m_3^2 > m_{3thr}^2$, no anomalous
threshold occurs.
\item If $m_3^2 < m_{3thr}^2$ the anomalous threshold has to be taken into
account. The behavior of its branch point $s_1$
at $m_{3thr}^2$ depends on the value of $\lambda(p_2^2,m_2^2,m_{3thr}^2)$.
\begin{itemize}
\item If $\lambda(p_2^2,m_2^2,m_{3thr}^2)<0$, which is equivalent to
$\lambda(p_3^2,m_1^2,m_{3thr}^2)<0$, the anomalous threshold deforms
the integration contour. Three subcases may occur:
\begin{itemize}
\item {\em Second case:} $s_1$ is real and $s_1<s_0$.
\item {\em Third case:} $s_1$ takes on complex values.
\item {\em Fourth case:} $s_1$ is real and $s_1>s_0$.
\end{itemize}
\item If $\lambda(p_2^2,m_2^2,m_{3thr}^2)>0$ the anomalous threshold deforms
the integration contour only for complex valued $s_1$. Two subcases can be
distinguished:
\begin{itemize}
\item {\em Fifth case:} $s_1$ is real and
$s_1>s_0$. Then the sheets of the logarithms in $\Delta C_0$ change.
\item {\em Sixth case:} $s_1$ takes on complex values.
\end{itemize}
\end{itemize}
\end{itemize}
For a numerical evaluation of the dispersion integral the following
procedure for handling these cases can be applied.
In the first case the discontinuity $\Delta C_0$
can be calculated as given in (\ref{DeltaC0}). In the other cases the
sheets of the logarithms in $\Delta C_0$ differ from (\ref{DeltaC0}).
In the second, third and sixth case the logarithms in (\ref{DeltaC0})
have to be replaced by
\begin{displaymath}
    \log(a+b+i\epsilon)-\log(a-b-i\epsilon) \, .
\end{displaymath}
In the fourth case the additional discontinuity of the logarithm yields
\begin{displaymath}
   \left\{ \begin{array}{ll}
           \log(a+b+i\epsilon)-\log(a-b-i\epsilon) - 2 \pi i
       & \quad \mbox{for} \quad s<s_1 \,,  \\
      \log(a+b+i\epsilon)-\log(a-b-i\epsilon)
     & \quad \mbox{for} \quad s>s_1 \, . \end{array} \right.
\end{displaymath}
In the fifth case the correct expressions are
\begin{displaymath}
   \left\{ \begin{array}{ll}
           \log(a+b-i\epsilon)-\log(a-b-i\epsilon)
       & \quad \mbox{for}\;s<s_1\;\mbox{and real}\;b\,,  \\
      \log(a+b)-\log(a-b) - 2 \pi i
     & \quad \mbox{for}\;s<s_1\;\mbox{and imaginary}\; b\,, \\
      \log(a+b+i\epsilon)-\log(a-b-i\epsilon)
      & \quad \mbox{for}\;s>s_1 \, .    \end{array}  \right.
\end{displaymath}

In the second, third and sixth case the contribution $C_{0an}$ of the
anomalous threshold, given by the expressions (\ref{C0an}) and
(\ref{C0an-2}), has to be added.
The argument of the
second logarithm in (\ref{C0an-2}) can get a negative imaginary part
if $s_1$ has a nonvanishing imaginary part, i.e.~in the third and sixth
case. It has then to be replaced by the appropriate analytic continuation,
which is given by
\begin{eqnarray}
   && -\log\big(-s_1(p_2^2+p_3^2-p_1^2)+(p_2^2-p_3^2)^2-
                       p_1^2(p_2^2+p_3^2) \nn
  \quad  \quad
               +  \sqrt{(\Sigma p)^2-s_1+i\epsilon}
                  \sqrt{(\Delta p)^2-s_1+i\epsilon}
                  \sqrt{\lambda(p_1^2,p_2^2,p_3^2)}
                \big) - 2 \pi i \, . \label{C0an-3}
\end{eqnarray}
The expressions (\ref{C0an-2}) and (\ref{C0an-3}) are valid for
$p_1^2>(|p_2|+|p_3|)^2$.
An analytic continuation in terms of $p_1^2$ is easy, but is not required
in the context of this paper.

We now return to the master diagram.
Formula (\ref{DT3}), which Broadhurst \cite{Broadhurst} took as starting point,
is particularly useful, because the square root belonging to
$\Delta B_0$ cancels one of the square roots of $\Delta C_0$,
which results in
\begin{equation}
  \DTM^{(3a)}(p^2;m_i^2)
     =  \frac{2 \pi i}{p^2}
    \int\limits_{(m_3+m_4)^2}^{(\sqrt{p^2}-m_2)^2} \!\! dt \,
      \frac{1}{t-m_1^2+i\epsilon} \,
      2 \, \tanh^{-1}\frac{b\,(t,m_2^2,p^2,m_4^2,m_3^2)}
                          {a\,(t,m_2^2,p^2,m_4^2,m_3^2,m_5^2)}
                    \, , \label{DT3-1}
\end{equation}
with $a$ and $b$ defined as in (\ref{a}) and (\ref{b}).
The argument of the $\tanh^{-1}$ is
a rational function of $t$, $p^2$, $m_2^2$, $m_3^2$,
$m_4^2$, $m_5^2$,
and of $\sqrt{\lambda(t,p^2,m_2^2) \lambda(t,m_3^2,m_4^2)}$.

\subsection{Representation of the three-particle discontinuities
in terms of complete logarithmic elliptic integrals}

A partial integration in (\ref{DT3-1}) yields for the three-particle
discontinuity
\begin{eqnarray}
  \DTM^{(3a)}(p^2;m_i^2) &=&
    - \frac{4 \pi i}{p^2}  \int\limits_{(m_3+m_4)^2}^{(\sqrt{p^2}-m_2)^2} dt
        \log \left( \frac{t}{m_1^2}-1 + i\epsilon \right)  \nonumber \\
  && \;\;\;\;\;\;\;\;\;\;\;\;\;\; \times
        \frac{\partial}{\partial t} \,
              \left( \tanh^{-1}
                                 \frac{b\,(t,m_2^2,p^2,m_4^2,m_3^2)}
                                      {a\,(t,m_2^2,p^2,m_4^2,m_3^2,m_5^2)}
                     \right)
             \, . \label{DT3par}
\end{eqnarray}

Except for the logarithm, the form of (\ref{DT3par}) equals those of
elliptic integrals.
The zeros of the square root
$\sqrt{\lambda(t,p^2,m_2^2) \lambda(t,m_3^2,m_4^2)}$ as a function
of $t$ are located at $(p \pm m_2)^2$ and $(m_3 \pm m_4)^2$.
Therefore we introduce in the following the characteristic variables
\begin{equation}
  q_{\pm \pm} = (p \pm m_2)^2 - (m_3 \pm m_4)^2 \, .
\end{equation}

In analogy to the case of the transformations of elliptic integrals
to the Legendre normal form \cite{Erd} one substitutes
\begin{equation}
   t = \frac{(m_3+m_4)^2 q_{--} - (m_3-m_4)^2 q_{-+} x^2}
              {q_{--} - q_{-+} x^2} \, ,
  \label{txsubst}
\end{equation}
which leads to
\begin{eqnarray}
 \DTM^{(3a)} &=& 2 \pi i
  \int\limits_0^1 \frac{dx} {\sqrt{(1-x^2)(1-\kappa^2 x^2)}}  \,\,
  \left\{ \log(a_0) + \log(1-a_1 x^2) - \log(1-a_2 x^2) \right\} \nonumber \\
  && \;\;\;\;\;\; \times  h^{(1)} \;
  \frac{r^{(1)}_4 q_{-+}^4 x^8 +  r^{(1)}_3 q_{-+}^3 q_{--} x^6
        + r^{(1)}_2 q_{-+}^2 q_{--}^2 x^4 + r^{(1)}_1 q_{-+} q_{--}^3 x^2
        + r^{(1)}_0 q_{--}^4}
       { \prod\limits_{i=1}^4 (q_{--}- c^{(1)}_i q_{-+} x^2) } \nonumber \\
  &=& 2 \pi i
  \int\limits_0^1 \frac{dx} {\sqrt{(1-x^2)(1-\kappa^2 x^2)}}  \,\,
  \left\{ \log(a_0) + \log(1-a_1 x^2) - \log(1-a_2 x^2) \right\} \nonumber \\
  && \;\;\;\;\;\; \times  h^{(1)}
  \left( h^{(1)}_0 + \sum_{i=1}^4 \frac{h^{(1)}_i}
                                 {1-c^{(1)}_i\frac{q_{-+}}{q_{--}} x^2}
    \right) \, , \label{imag1}
\end{eqnarray}
with
\begin{eqnarray}
 && a_0 = \frac{(m_3+m_4)^2}{m_1^2} \, , \quad
    a_1 = \frac{(m_3-m_4)^2 -m_1^2}{(m_3+m_4)^2 -m_1^2}
         \frac{q_{-+}}{q_{--}} \, , \quad
    a_2 = \frac{q_{-+}}{q_{--}} \, , \nn
   h^{(1)} =  \frac{4 \, m_3 m_4}
                   {p^2 \, \sqrt{q_{++}q_{--}} \, (m_3+m_4)^2
                    \left[m_3(p^2-m_4^2-m_5^2)+m_4(m_2^2-m_3^2-m_5^2
                    \right]^2} \, , \nn
   h^{(1)}_0=\frac{r^{(1)}_4}{c^{(1)}_1 c^{(1)}_2 c^{(1)}_3 c^{(1)}_4} \, , \nn
   h^{(1)}_i=\frac{r^{(1)}_4 + r^{(1)}_3 c^{(1)}_i
                   + r^{(1)}_2 (c^{(1)}_i)^2 + r^{(1)}_1 (c^{(1)}_i)^3
                   + r^{(1)}_0 (c^{(1)}_i)^4}
            {c^{(1)}_i \prod\limits_{j \neq i} (c^{(1)}_i - c^{(1)}_j)} \quad
          \mbox{for $i=1,\ldots,4$} \, , \nn
   c^{(1)}_1=1 \, , \quad
   c^{(1)}_2=\frac{(m_3-m_4)^2}{(m_3+m_4)^2} \, , \nn
   c^{(1)}_{3/4} =
     \frac{ (p^2-m_4^2-m_5^2)(m_2^2-m_3^2-m_5^2) - 4 m_3 m_4 m_5^2 \pm
                    \sqrt{\lambda(m_5^2,p^2,m_4^2) \lambda(m_5^2,m_2^2,m_3^2)}}
        { (p^2-m_4^2-m_5^2)(m_2^2-m_3^2-m_5^2) + 4 m_3 m_4 m_5^2 \pm
                    \sqrt{\lambda(m_5^2,p^2,m_4^2) \lambda(m_5^2,m_2^2,m_3^2)}}
   \, , \nn
  r^{(1)}_0 =  (m_3 + m_4) q_{-+} q_{++}
  (m_2^2 m_4 - m_3^2 m_4 - m_3 m_4^2 - m_3 m_5^2 - m_4 m_5^2 + m_3 p^2) \,, \nn
  r^{(1)}_1 = -2  \Big\{ \big[
                       -m_4^2 (p^2 -m_2^2)^3 \nn
           \;\;\;\;\;\;\;\;\;\;\;\;\;\;\;       +2 \, m_4 (p^2 -m_2^2)
                         (-m_2^4 m_3 + 2 m_2^2 m_3 m_4^2 + 3 m_2^2 m_4^3 \nn
           \;\;\;\;\;\;\;\;\;\;\;\;\;\;\;\;\;\;\;\;\;\;\;\;\;\;\;\;\;\;
           \;\;\;\;\;\;\;\;\;\;
                          + m_2^2 m_3 p^2 + 6 m_3 m_4^2 p^2 + 5 m_4^3 p^2) \nn
           \;\;\;\;\;\;\;\;\;\;
           \;\;\;\;\;  +(m_3+m_4)^2
                   (-6 m_2^4 m_3 m_4 + 6 m_2^2 m_3^3 m_4 - 2 m_3^5 m_4
                    + 4 m_2^4 m_4^2 \nn
           \;\;\;\;\;\;\;\;\;\;
           \;\;\;\;\;\;\;\;\;\;\;\;\;\;\;\;\;\;\;\;\;\;\;\;\;\;
                    + m_2^2 m_3^2 m_4^2 - 2 m_3^4 m_4^2
                    + m_2^2 m_4^4 - 2 m_3^2 m_4^4 \nn
           \;\;\;\;\;\;\;\;\;\;
           \;\;\;\;\;\;\;\;\;\;\;\;\;\;\;\;\;\;\;\;\;\;\;\;\;\;
                    - 2 m_3 m_4^5
                    + m_2^4 p^2 -  2 m_2^2 m_3^2 p^2 + m_3^4 p^2  \nn
           \;\;\;\;\;\;\;\;\;\;
           \;\;\;\;\;\;\;\;\;\;\;\;\;\;\;\;\;\;\;\;\;\;\;\;\;\;
                    + 4 m_2^2 m_3 m_4 p^2 + 2 m_2^2 m_4^2 p^2
                    +  m_3^2 m_4^2 p^2 + 6 m_3 m_4^3 p^2 \nn
           \;\;\;\;\;\;\;\;\;\;
           \;\;\;\;\;\;\;\;\;\;\;\;\;\;\;\;\;\;\;\;\;\;\;\;\;\;
                    - 2 m_2^2 p^4
                    - 2 m_3^2 p^4 + 2 m_3 m_4 p^4 - 10 m_4^2 p^4 + p^6)
                   \big] \nn
           \;\;\;\;\;\;\;\;\;\;
           \;\;\;\;\; +  m_5^2 \big[
                       -2 m_3 m_4 (p^2-m_2^2)^2 \nn
           \;\;\;\;\;\;\;\;\;\;
           \;\;\;\;\;\;\;\;\;\;\;\;\;
                  +(m_3+m_4)^2
                   (-(p^2-m_2^2)^2 + 2 m_2^2 m_3^2 - m_3^4
                    + 2 m_3^3 m_4 \nn
           \;\;\;\;\;\;\;\;\;\;
           \;\;\;\;\;\;\;\;\;\;\;\;\;\;\;\;\;\;\;\;\;\;\;\;\;\;\;\;\;\;\;\;\;\;
                    + 2 m_2^2 m_4^2 + 6 m_3^2 m_4^2 + 2 m_3 m_4^3 \nn
           \;\;\;\;\;\;\;\;\;\;
           \;\;\;\;\;\;\;\;\;\;\;\;\;\;\;\;\;\;\;\;\;\;\;\;\;\;\;\;\;\;\;\;\;\;
                    - m_4^4 + 2 m_3^2 p^2 + 2 m_4^2 p^2 )
                   \big]
           \Big\} \, , \nn
 r^{(1)}_2 = 6\, m_3 m_4 (p^2-m_2^2 + m_3^2 - m_4^2)
              (\lambda(p^2,m_4^2,m_5^2)-\lambda(m_2^2,m_3^2,m_5^2)) \,, \nn
 r^{(1)}_3 = -2  \Big\{ \big[
                    m_4 (p^2 -m_2^2)
           (-2 m_2^4 m_3 + m_2^4 m_4 + 4 m_2^2 m_3 m_4^2
            - 6 m_2^2 m_4^3
            + 2 m_2^2 m_3 p^2 \nn
      \;\;\;\;\;\;\;\;\;\;\;\;\;\;\;\;\;\;\;\;
      \;\;\;\;\;\;\;\;\;\;\;\;\;\;\;\;\;\;
            - 2 m_2^2 m_4 p^2 + 12 m_3 m_4^2 p^2
            - 10 m_4^3 p^2 + m_4 p^4 ) \nn
           \;\;\;\;\;\;\;\;\;\;
               \;\;\;\;\; +(m_3-m_4)^2
                     (-6 m_2^4 m_3 m_4 + 6 m_2^2 m_3^3 m_4
                     - 2 m_3^5 m_4  + 10 m_4^2 p^4 - p^6 \nn
           \;\;\;\;\;\;\;\;\;\;
           \;\;\;\;\;\;\;\;\;\;\;\;\;\;\;\;\;\;\;\;\;\;\;\;\;\;
                     - 4 m_2^4 m_4^2 - m_2^2 m_3^2 m_4^2
                     + 2 m_3^4 m_4^2 - m_2^2 m_4^4 \nn
           \;\;\;\;\;\;\;\;\;\;
           \;\;\;\;\;\;\;\;\;\;\;\;\;\;\;\;\;\;\;\;\;\;\;\;\;\;
                     + 2 m_3^2 m_4^4
                     - 2 m_3 m_4^5 - m_2^4 p^2 +  2 m_2^2 m_3^2 p^2  \nn
           \;\;\;\;\;\;\;\;\;\;
           \;\;\;\;\;\;\;\;\;\;\;\;\;\;\;\;\;\;\;\;\;\;\;\;\;\;
                     - m_3^4 p^2 + 4 m_2^2 m_3 m_4 p^2
                     - 2 m_2^2 m_4^2 p^2 -   m_3^2 m_4^2 p^2 \nn
           \;\;\;\;\;\;\;\;\;\;
           \;\;\;\;\;\;\;\;\;\;\;\;\;\;\;\;\;\;\;\;\;\;\;\;\;\;
                     + 6 m_3 m_4^3 p^2 + 2 m_2^2 p^4 + 2 m_3^2 p^4
                     + 2 m_3 m_4 p^4 ) \big] \nn
           \;\;\;\;\;\;\;\;\;\;
           \;\;\;\;\;
             +m_5^2 \big[
                  -2 m_3 m_4 (p^2-m_2^2)^2 \nn
           \;\;\;\;\;\;\;\;\;\;
           \;\;\;\;\;\;\;\;\;\;\;\;\;
                  +(m_3-m_4)^2
                   ((p^2-m_2^2)^2 -2 m_2^2 m_3^2 + m_3^4
                    + 2 m_3^3 m_4 \nn
           \;\;\;\;\;\;\;\;\;\;
           \;\;\;\;\;\;\;\;\;\;\;\;\;\;\;\;\;\;\;\;\;\;\;\;\;\;\;\;\;\;\;\;\;\;
                    - 2 m_2^2 m_4^2
                    - 6 m_3^2 m_4^2 + 2 m_3 m_4^3  \nn
           \;\;\;\;\;\;\;\;\;\;
           \;\;\;\;\;\;\;\;\;\;\;\;\;\;\;\;\;\;\;\;\;\;\;\;\;\;\;\;\;\;\;\;\;\;
                    +  m_4^4 - 2 m_3^2 p^2 - 2 m_4^2 p^2 )
                   \big]
           \Big\} \, , \nn
 r^{(1)}_4 = q_{--} q_{+-} (m_4 - m_3)
       \left[ m_3(p^2-m_4^2-m_5^2) - m_4 (m_2^2-m_3^2-m_5^2) \right] \, .
  \nonumber
\end{eqnarray}
The integrals in (\ref{imag1}) are either of the type of
complete elliptic integrals or they have the structure
\begin{eqnarray}
 {\rm LK} (a,\kappa) &=&
  \int\limits_0^1 \frac{\log(1-a x^2) dx}{\sqrt{(1-x^2)(1-\kappa^2 x^2)}}
           \;\, \label{LEK} \\
 {\rm L} \Pi (a,c,\kappa) &=&
  \int\limits_0^1 \frac{\log(1-a x^2) dx}{\sqrt{(1-x^2)(1-\kappa^2 x^2)}}
        \frac{1}
             {1-c x^2} \, . \label{LEP}
\end{eqnarray}
The analogy of these functions to the elliptic integrals is obvious.
According to our knowledge they have not been introduced
in the mathematical
literature. We call them complete logarithmic elliptic integrals.
For sake of completeness one can introduce
a third function,
\begin{eqnarray}
 {\rm LE} (a,\kappa) &=& \int\limits_0^1 \frac{(1-\kappa^2 x^2) \log(1-a x^2)
dx}
                     {\sqrt{(1-x^2)(1-\kappa^2 x^2)}} \, . \label{LEE}
\end{eqnarray}
Then all integrations of the type
\begin{eqnarray}
    \int\limits_0^1 \frac{\log(1-a x^2) dx}
                     {\sqrt{(1-x^2)(1-\kappa^2 x^2)}} \,
           r(x) \, , \label{EL}
\end{eqnarray}
where $r(x)$ is an arbitrary rational function, can be expressed in terms of
${\rm LK}$, ${\rm LE}$, ${\rm L}\Pi$, complete elliptic integrals,
logarithms and polylogarithms.

The functions ${\rm LK}$, ${\rm LE}$ and ${\rm L}\Pi$ are related to
generalized hypergeometric functions of two and three variables, i.e.~the
Appell function $F_1(\alpha;\beta_1,\beta_2;\gamma;x,y)$ and the
Lauricella function $F_D^{(3)}(\alpha;\beta_1,\beta_2,\beta_3;\gamma;x,y)$
\cite{hypergeometric},
\begin{eqnarray}
 {\rm LK} (a,\kappa) &=& - \frac{\pi}{2}
     \frac{\partial}{\partial \beta_2}      \left.
             \rmF_1(\frac{1}{2};\frac{1}{2},\beta_2;1;\kappa^2,a)
            \right| _{\beta_2=0} \,, \\
 {\rm LE} (a,\kappa) &=& - \frac{\pi}{2}
     \frac{\partial}{\partial \beta_2}      \left.
            \rmF_1(\frac{1}{2};-\frac{1}{2},\beta_2;1;\kappa^2,a)
            \right| _{\beta_2=0} \,, \\
  {\rm L} \Pi (a,c,\kappa) &=& - \frac{\pi}{2}
     \frac{\partial}{\partial \beta_3}      \left.
           {\rm F}_{\rm D}(\frac{1}{2};\frac{1}{2},1,\beta_3;1;\kappa^2,c,a)
            \right| _{\beta_3=0} \,.
\end{eqnarray}
The representation in terms of generalized hypergeometric functions
helps to perform analytic continuations
for ${\rm LK}$, ${\rm LK}$ and ${\rm L} \Pi$. Series representations
are given as
\begin{eqnarray}
 {\rm LK} (a,\kappa) &=&
  - \frac{\pi}{4} \, a \sum_{m,n=0}^\infty
      \frac{(\frac{1}{2})_m(\frac{3}{2})_{m+n}}{m!(m+n+1)!(n+1)}
      (\kappa^2)^m a^n \, , \label{LEKa} \\
 {\rm LE} (a,\kappa) &=&
  - \frac{\pi}{4} \, a \sum_{m,n=0}^\infty
      \frac{(-\frac{1}{2})_m (\frac{3}{2})_{m+n}}{m!(m+n+1)!(n+1)}
      (\kappa^2)^m a^n \, , \label{LEEa} \\
  {\rm L} \Pi (a,c,\kappa) &=&
  - \frac{\pi}{4} \, a \sum_{m,n,l=0}^\infty
      \frac{(\frac{1}{2})_m (\frac{3}{2})_{m+n+l}}{m!(m+n+l+1)!(l+1)}
      (\kappa^2)^m c^n a^l \, . \label{LEPa}
\end{eqnarray}
The region of convergence is given by $|\kappa^2|<1$, $|a|<1$, $|c|<1$.

In terms of these functions one can finally express the result for
$\DTM^{(3a)}$ as
\begin{eqnarray}
 \DTM^{(3a)}(p^2;m_i^2) &=&
      2 \pi i \, h^{(1)} \, \Bigg\{
            \;\;\log(a_0)
            \left( h^{(1)}_0 {\rm K} (\kappa)
                + \sum_{i=1}^4 h^{(1)}_i \,
                       \Pi(\frac{q_{-+}}{q_{--}} c^{(1)}_i,\kappa)
            \right) \nonumber \\
      && \;\;\;\;\;\;
        + \left( h^{(1)}_0 {\rm LK} (a_1,\kappa)
           + \sum_{i=1}^4 h^{(1)}_i \,
              {\rm L} \Pi(a_1,\frac{q_{-+}}{q_{--}} c^{(1)}_i,\kappa)
         \right) \nonumber \\
      && \;\;\;\;\;\;
         -\left( h^{(1)}_0 {\rm LK} (a_2,\kappa)
           + \sum_{i=1}^4 h^{(1)}_i \,
              {\rm L} \Pi(a_2,\frac{q_{-+}}{q_{--}} c^{(1)}_i,\kappa)
         \right) \Bigg\} \, , \;\;  \label{DT3-series}
\end{eqnarray}
a result which involves in an analogous way complete elliptic integrals
and the complete logarithmic integrals introduced in (\ref{LEK}),
(\ref{LEP}) and (\ref{LEE}).

\section{One-dimensional integral representations}
\label{one-dimensional-representations}

In many cases dispersion representations
have proven to be useful for the calculation of
two-loop self-energy integrals \cite{Broadhurst,ScharfTausk}.
One can use this approach for further analytical calculations, or
evaluate numerically the dispersion integral.
In the case of the master diagram, the discontinuity in the general
mass case is complicated to calculate, as shown in the last section.
We will present two ways to
simplify the evaluation of the dispersion integral.
The first one is a realization of an idea of Broadhurst \cite{Broadhurst}.
It  simplifies the evaluation of the three-particle cut contributions by
a partial integration in the dispersion integral.
The second way relies on interchanging the integrations in both the
two-particle cut contributions and the three-particle cut contributions
of the master diagram
and leads to a one-dimensional integral representation involving only
elementary functions.

\boldmath
\subsection{Evaluation of the three-particle cut contribution}
\unboldmath

Formula (\ref{DT2a}) expresses the two-particle cut contributions to
the discontinuity in terms of well-known functions, though involving
dilogarithms.
However, for the three-particle cuts, equation (\ref{DT3})
gives only an integral representation and expression (\ref{DT3-series})
is a solution in terms of rather complicated series representations.

As Broadhurst has pointed out \cite{Broadhurst}, it is useful to perform
a partial integration in the dispersion integral for the
three-particle cut contribution.
One is then led to a suitable derivative, $\sigma'(p^2;m_i^2)$, of
the discontinuity,
\begin{equation}
 \sigma'(p^2;m_i^2) =
  \frac{1}{2 \pi i} \frac{\partial}{\partial p^2}
     \left( p^2 \Delta T(p^2;m_i^2) \right)
  = \frac{1}{2 \pi i} \left(
 \Delta T(p^2;m_i^2)
     + p^2 \frac{\partial \Delta T(p^2;m_i^2)}{\partial p^2}
           \right)  \, .
\end{equation}

The function $T$ can then be calculated from $\sigma'$ as
\begin{equation}
   T(p^2;m_i^2) =
         - \frac{1}{p^2} \int\limits_{s_0} ^\infty ds \,
               \log(s-p^2-i\epsilon) \, \sigma' (s;m_i^2) \; .
    \label{T3-sigma}
\end{equation}

Equation (\ref{DT3-1}) yields for $\sigma^{(3a)'}(p^2;m_i^2)$
of the three-particle cut contribution $\TM^{(3a)}$
\begin{equation}
  \sigma^{(3a)'}(p^2;m_i^2)
           =  2 \int\limits_{(m_3+m_4)^2}
                        ^{(\sqrt{p^2}-m_2)^2} \frac{dt}{t-m_1^2}
                    \frac{\partial}
                         {\partial p^2} \,
                              \left(\tanh^{-1}
                     \frac{b\,(t,m_2^2,p^2,m_4^2,m_3^2)}
                          {a\,(t,m_2^2,p^2,m_4^2,m_3^2,m_5^2)}
                               \right) \, .
\end{equation}
After a lengthy decomposition into partial fractions
one finally arrives at a
surprisingly simple result, involving only three $\Pi$-functions,
i.e.~complete elliptic integrals of the third kind,
\begin{eqnarray}
  \sigma^{(3a)'}(p^2;m_i^2) &=&
        \int\limits_0^1 \frac{dx}{\sqrt{(1-x^2)(1-\kappa^2 x^2)}}
        h^{(2)}  \sum_{i=1}^3
           \frac{r^{(2)}_1 q_{--} q_{-+}^2 x^4
                           + r^{(2)}_0 q_{--}^2 q_{-+} x^2}
                          {\prod\limits_{i=1}^3
                              (q_{--} - q_{-+} c^{(2)}_i x^2)}
                \nonumber \\
         &=&
        \int\limits_0^1 \frac{dx}{\sqrt{(1-x^2)(1-\kappa^2 x^2)}}
        h^{(2)} \sum_{i=1}^3 \frac{h^{(2)}_i}
                                  {1-\frac{q_{-+}}{q_{--}} c^{(2)}_i x^2}
                \nonumber \\
         &=&
 h^{(2)} \sum_{i=1}^3
        h_i^{(2)} \Pi(\frac{q_{-+}}{q_{--}} c_i^{(2)},\kappa)
  \, ,
    \label{Dsigma}
\end{eqnarray}
with variables,
\begin{eqnarray}
  \kappa^2 &=&\frac{q_{+-} q_{-+}}
                   {q_{++} q_{--}} \, , \quad
  c_1^{(2)} \;=\; \frac{(m_3-m_4)^2-m_1^2}{(m_3+m_4)^2-m_1^2} \, ,
  \nonumber \\
  c_{2/3}^{(2)} &=&
     \frac{ (p^2-m_4^2-m_5^2)(m_2^2-m_3^2-m_5^2) - 4 m_3 m_4 m_5^2 \pm
                    \sqrt{\lambda(m_5^2,p^2,m_4^2) \lambda(m_5^2,m_2^2,m_3^2)}}
        { (p^2-m_4^2-m_5^2)(m_2^2-m_3^2-m_5^2) + 4 m_3 m_4 m_5^2 \pm
                    \sqrt{\lambda(m_5^2,p^2,m_4^2) \lambda(m_5^2,m_2^2,m_3^2)}}
   \, , \nonumber \\
 h^{(2)} &=&  \frac{32 m_3^2 m_4^2}
                {\sqrt{q_{++} q_{--}}
                \left[m_3(p^2-m_4^2-m_5^2)+m_4(m_2^2-m_3^2-m_5^2)
                         \right]^2 \left[(m_3+m_4)^2-m_1^2\right]} \, ,
                \nonumber \\
 h_i^{(2)} &=& \frac{r^{(2)}_1+r^{(2)}_0 c_i^{(2)}}
              {\prod\limits_{j \neq i} (c_i^{(2)}-c_j^{(2)})}
     \quad \mbox{for $i=1,2,3$} \, ,
  \nonumber \\
  r^{(2)}_0
    &=& (m_2^2-m_3^2)^2 - 2 m_3 m_4 (m_2^2 - m_3^2) + m_4^2 (m_2^2+m_3^2)
       \nonumber \\
  &&  - m_5^2 (m_2^2 + (m_3+m_4)^2) - p^2 (m_2^2 + m_3^2 - m_5^2)
   \, ,\nonumber \\
  r^{(2)}_1 &=& - r^{(2)}_0 ( m_4 \leftrightarrow -m_4) \, .
\nonumber
\end{eqnarray}

\subsection{An efficient one-dimensional integral representation
for the master diagram}

The dispersion integral for the master diagram involves in the integrand
dilogarithms for the two-particle cut contributions
and complete elliptic integrals for the three-particle cut contributions
(\ref{T3-sigma}), as presented in the previous section.
In this section we use a different approach in order
to derive a one-dimensional integral representation with
elementary functions. This is achieved in the following way.

For an evaluation of the two-particle cut contributions
it proves useful to introduce the dispersion
representation of the function $C_0$ as given in equation (\ref{dispC0}).
After an insertion of (\ref{DT2a}) and (\ref{dispC0})
into (\ref{T-contrib-2a}), the integrations in the
dispersion integral can be interchanged and lead to
\begin{eqnarray}
  &&
      \TM^{(2a)}(p^2;m_1^2,m_2^2,m_3^2,m_4^2,m_5^2) \nonumber \\
  && \;\;\;= \frac{1}{2 \pi i} \int\limits_{s_0}^\infty
           \frac{ds}{s-p^2-i\epsilon}
           \DTM^{(2a)}(s;m_i^2) \nonumber \\
  && \;\;\;= - \left( \frac{1}{2 \pi i} \right)^2 \int\limits_{t_0}^\infty dt
        \int\limits_{s_0}^\infty ds
        \frac{\left( \Delta C_0(t) \right)^* \Delta B_0 (s)}
             {(s-p^2-i\epsilon )(t-s + i\epsilon )}
        \nonumber \\
  && \;\;\;\;\;\; + \frac{1}{2\pi i} \int\limits_{s_0}^\infty ds \,
               \frac{\Delta B_0(s;m_1^2,m_2^2)}
                    {s-p^2-i\epsilon}
          \left( C_{0an} (s,m_1^2,m_2^2;m_5^2,m_4^2,m_3^2) \right)^*
      \nonumber \\
  && \;\;\;
     =  - \frac{1}{2 \pi i} \int\limits_{(m_4+m_5)^2}^\infty \!\!\!\! dt
        \frac{ \left(
               \Delta C_0 (t,m_1^2,m_2^2; m_5^2,m_4^2, m_3^2)
               \right) ^*  }
              {t-p^2-i\epsilon}
        \left(B_0(p^2;m_1^2,m_2^2) - B_0 (t;m_1^2,m_2^2) \right)
      \nonumber \\
  &&     \;\;\;\;\;\;
      + \frac{1}{2\pi i} \int\limits_{(m_1+m_2)^2}^\infty ds \,
               \frac{\Delta B_0(s;m_1^2,m_2^2)}
                    {s-p^2-i\epsilon}
        \left( C_{0an} (s,m_1^2,m_2^2;m_5^2,m_4^2,m_3^2) \right) ^* \, .
    \label{T2}
\end{eqnarray}
Thus, one ends up with two one-dimensional integrals. The infinite and
the constant parts of the two $B_0$-functions \cite{tHooft2}
in (\ref{T2}) cancel.
In both integrals the
integrands are composed of logarithms and square roots only,
as can be seen from equations (\ref{DeltaC0}) and (\ref{C0an-2}).
The two integrals can therefore easily be evaluated numerically.

For $\DTM^{(2b)}$ the analogous calculation yields
\begin{eqnarray}
  &&
      T^{(2b)}(p^2;m_1^2,m_2^2,m_3^2,m_4^2,m_5^2) \nonumber \\
  && \;\;\;
     =  \frac{1}{2 \pi i} \int\limits_{(m_1+m_2)^2}^\infty \!\!\!\! dt
        \frac{
               \Delta C_0 (t,m_4^2,m_5^2; m_2^2,m_1^2, m_3^2)
                 }
              {t-p^2-i\epsilon}
        \left(B_0(p^2;m_4^2,m_5^2) -
                \left( B_0 (t;m_4^2,m_5^2) \right)^* \right)
      \nonumber \\
  &&     \;\;\;\;\;\;
      + \frac{1}{2\pi i} \int\limits_{(m_4+m_5)^2}^\infty ds \,
               \frac{\Delta B_0(s;m_4^2,m_5^2)}
                    {s-p^2-i\epsilon}
         C_{0an} (s,m_4^2,m_5^2;m_2^2,m_1^2,m_3^2)  \, .
    \label{T2b}
\end{eqnarray}

The three-particle discontinuity (\ref{DT3}) can be transformed with
the partial integration (\ref{DT3par}), yielding
\begin{eqnarray}
  \DTM^{(3a)}(s;m_i^2) &=& 2 \pi i
     \int\limits_{(m_3+m_4)^2}^{(\sqrt{s}-m_2)^2} \!\! dt \,
      \log \left( \frac{t}{m_1^2}-1+i\epsilon \right) \nonumber \\
 &&  \;\;\;\;\;\;\;\;\;\;\;\;\;\;
     \times  \frac{ \sqrt{\lambda(s,t,m_2^2)} \,
       R(s,t,m_2^2,m_3^2,m_4^2,m_5^2) }{s}
               \label{T3partint} \, ,
\end{eqnarray}
where $R(s)$ is a rational function of s.
In the general mass case, i.e.~if all masses are different,
it has only first order poles.
This form is well suited for evaluating the dispersion integral,
\begin{equation}
   \TM^{(3a)}(p^2;m_i^2) =
 \frac{1}{2 \pi i}
 \int\limits_{s_0}^\infty ds \,
    \frac{\DTM^{(3a)} (s;m_i^2)}
         {s-p^2-i\epsilon}  \, ,
\end{equation}
because $R(s)$ can be decomposed into partial fractions,
\begin{eqnarray}
  R(s) &\!\!=&\!\!
   - \frac{\sqrt{\lambda(t,m_3^2,m_4^2)}}{t\, m_3^2} \frac{1}{(s-s_1)(s-s_2)}
     \left( \frac{r^{(3)}_1 s + r^{(3)}_2}{(s-s_3)(s-s_4)} + r^{(3)}_3 s +
r^{(3)}_4 \right)
       \nonumber \\
  &\!\! =&\!\! - \frac{\sqrt{\lambda(t,m_3^2,m_4^2)}}{t\, m_3^2}
                \sum_{i=1}^4 \frac{f_i(t,m_2^2,m_3^2,m_4^2,m_5^2)}{s-s_i} \,.
\end{eqnarray}
The values of the $r^{(3)}_i$ and $s_i$ will be presented below.
An interchange of integrations,
\begin{equation}
   \int\limits_{(m_2+m_3+m_4)^2}^\infty ds
        \int\limits_{(m_3+m_4)^2}^{(\sqrt{s}-m_2)^2} dt
   =  \int\limits_{(m_3+m_4)^2}^\infty dt
     \int\limits_{(m_2+\sqrt{t})^2}^\infty ds \,,
\end{equation}
leads to integrations of the type
\begin{equation}
   \int\limits _{(\sqrt{t}+m_2)^2}^\infty \frac{ds}{s-s_i}
             \frac{\sqrt{\lambda(s,t,m_2^2)}}{s}
     =  B_0(s_i;t,m_2^2) \,,
\end{equation}
i.e.~one-loop self-energy integrals. In the final expression all
ultraviolet divergencies of these $B_0$-functions cancel.
The result for the three-particle cut contribution of the master
diagram is then
\begin{eqnarray}
  && \!\!\!\!\!  T^{(3a)}(p^2;m_i^2) =
        \int\limits_{(m_3+m_4)^2}^\infty dt \;
        \log \left( \frac{t}{m_1^2}-1+i\epsilon \right)
     \nonumber \\
  && \hskip40mm \times
        \frac{\sqrt{\lambda(t,m_4^2,m_3^2)}}{t\, m_3^2} \;
        \sum_{i=1}^4 f_i
        \frac{B_0(p^2;t,m_2^2)-B_0(s_i;t,m_2^2)} {s_i-p^2-i\epsilon} \,,
       \label{T12345-3} \\
  && \!\!\!\!\! \mbox{with} \nonumber \\
  &&\!\!\!\!\! s_{1/2}=
           \frac{t+m_2^2+m_4^2+m_5^2-m_3^2}{2}
          +   \frac{(t-m_4^2)(m_5^2-m_2^2) \pm
                   \sqrt{\lambda(t,m_3^2,m_4^2)\lambda(m_2^2,m_3^2,m_5^2)}}
                   {2m_3^2}
                 \nn
   \!\!\!\!\! s_{3/4}=(m_2 \pm \sqrt{t})^2 \,, \nn
   \!\!\!\!\! f_1=
    \frac{s_1 r^{(3)}_1 +r^{(3)}_2}{\prod\limits_{i=2}^4 (s_1-s_i)}
                 + \frac{s_1 r^{(3)}_3 + r^{(3)}_4}{s_1-s_2} \,,
      \;\; f_2=f_1(s_1 \leftrightarrow s_2) \,, \nonumber\\
  && \!\!\!\!\! f_3=\frac{s_3 r^{(3)}_1 + r^{(3)}_2}
                         {\prod\limits _{i\neq 3} (s_3-s_i)} \,,
      \;\;
     f_4=f_3(s_3 \leftrightarrow s_4) \,, \nn
   \!\!\!\!\! r^{(3)}_1=t(2m_2^2-m_5^2+m_3^2)-m_2^2(m_4^2-m_3^2) \,, \nn
   \!\!\!\!\! r^{(3)}_2=(t-m_2^2)(t(m_5^2-m_3^2)-m_2^2(m_4^2-m_3^2)) \,, \nn
   \!\!\!\!\! r^{(3)}_3=\frac{m_3^2(m_3^2-t-m_4^2)}
                         {\lambda(t,m_4^2,m_3^2)} \,, \nn
   \!\!\!\!\! r^{(3)}_4=m_3^2 \nn
   \;     + \frac{t(2m_4^2m_3^2-m_4^2m_2^2+m_5^2m_4^2+m_5^2m_3^2)
               + m_4^2(m_2^2-m_5^2)(m_4^2-m_3^2)
               + m_5^2 m_3^2 (m_4^2-m_3^2)}
              {\lambda(t,m_4^2,m_3^2)} \, . \nonumber
\end{eqnarray}

In order to obtain the
second three-particle cut contribution $T^{(3b)}$, the masses have to
be interchanged according to
$m_1 \leftrightarrow m_2$ and $m_4 \leftrightarrow m_5$.

In some parameter regions one has to be careful when
evaluating the $B_0$-functions in (\ref{T12345-3}).
The external momentum variable $p^2$ can be assumed
to have an infinitely small positive
imaginary part. Consequently, $B_0(p^2+\ieps;t,m_2^2)$
can be evaluated in the usual way.
The functions $B_0(s_{3/4};t,m_2^2)$ are always evaluated below the
threshold, and present therefore no difficulty.
However, care has to be taken in the evaluation of $B_0(s_{1/2};t,m_2^2)$.
A brief inspection of the asymptotic behavior of $s_{1/2}$ shows
that the functions
$B_0(s_{1/2};t,m_2^2)$ have to be evaluated below the threshold for large
values of $m_3^2$. In fact, no difficulty occurs for $m_3^2>(m_2+m_5)^2$.
But $s_1$ and $s_2$ get an imaginary part for
\begin{displaymath}
  (m_2-m_5)^2 < m_3^2 < (m_2+m_5)^2 \, .
\end{displaymath}
In that case, the functions $B_0(s_{1/2};m_1^2,m_2^2)$
have to be evaluated for complex values of the
variables $s_{1/2}$, which is not too complicated.
For  $m_3^2<(m_2-m_5)^2$, the imaginary part of the variables $s_{1/2}$
vanishes again. Then one can in general evaluate the $B_0$-functions
with the usual algorithm which assumes
$B_0(s_{1/2}+i\epsilon;t,m_2^2)$. However, there exists a threshold
$t_{thr1}$,
\begin{eqnarray}
  t_{thr1} &=& \frac{1}{2 m_2^2}
               \bigg( \lambda(m_3^2,m_2^2,m_5^2) + 2 m_2^2 (m_3^2+m_4^2)
       \nonumber \\
 && \quad \quad \quad \quad
                    +(m_5^2-m_2^2-m_3^2)
                      \sqrt{\lambda(m_2^2,m_3^2,m_5^2)+4 m_2^2 m_4^2}
                           \bigg) \, , \label{limit1}
\end{eqnarray}
and for $t>t_{thr1}$ the function $B_0(s_2;t,m_2^2)$ has to be evaluated as
$B_0(s_2-i\epsilon;t,m_2^2)=(B_0(s_2+i\epsilon;t,m_2^2))^*$.

Another complication occurs in that case,
if $s_2(t)$ equals $p^2$ at a value $t>t_{thr1}$. Then one denominator
in (\ref{T12345-3}) vanishes at a point where the numerator takes on the
value of the discontinuity of a $B_0$-function.
This happens at $t=t_{thr2}$, if the following conditions apply
\begin{eqnarray}
  && \lambda(p^2,m_4^2,m_5^2) > 0 \label{lim2-1} \,, \\
  && t_{thr1} < t_{thr2} =
   \frac{1}{2 m_5^2} \bigg(
         m_5^2(p^2+m_2^2+m_3^2+m_4^2-m_5^2) + (m_2^2-m_3^2)(m_4^2-p^2)
      \nonumber \\
 &&\quad \quad \quad\quad \quad \quad\quad\quad\quad
    + \sqrt{\lambda(m_2^2,m_3^2,m_5^2) \lambda(p^2,m_4^2,m_5^2)}
          \bigg) \label{lim2-2}  \,, \\
  && t_{thr2} > t_0 = (m_3+m_4)^2 \label{lim2-3} \,, \\
  && m_3^2+m_5^2-m_2^2-\sqrt{\lambda(m_2^2,m_3^2,m_5^2)} > 0 \, .
                 \label{lim2-4}
\end{eqnarray}
The condition (\ref{lim2-1}) is necessary for $t_{thr2}$ to be
real valued.
The conditions (\ref{lim2-2}) and (\ref{lim2-3}) are obviously necessary
for the complication in question. The condition (\ref{lim2-4})
can be understood from the asymptotic behavior of $s_2$ for large values
of $t$.
Thus, if $m_3^2<(m_2-m_5)^2$ and the conditions
(\ref{lim2-1}-\ref{lim2-4}) apply,
the principal value and the residue of the integral have to be calculated.

The calculation of the residue involves the derivative
of $s_2$ with respect to $t$,
\begin{equation}
  \frac{ds_2}{dt} = \frac{1}{2 \, m_3^2}
        \left( m_3^2 -m_2^2+m_5^2
                          + (m_3^2+m_4^2-t)
                        \sqrt{\frac{\lambda(m_2^2,m_3^2,m_5^2)}
                                   {\lambda(t,m_3^2,m_4^2)}} \right)
                     \, .
\end{equation}
The calculation then yields
\begin{eqnarray}
 &&  \int\limits_{t_0}^\infty dt \, f(t)
      \frac{B_0(p^2;t,m_2^2)-B_0(s_2(t);t,m_2^2)}{s_2-p^2-i\epsilon}  \\
 && \quad \quad \quad \quad = \mbox{\large  P}
     \int\limits_{t_0}^\infty dt \, f(t)
       \frac{B_0(p^2;t,m_2^2)-B_0(s_2(t);t,m_2^2)}{s_2-p^2} \\
 && \quad \quad \quad \quad \quad
   + i \pi f(t_{thr2}) \frac{ \Delta B_0(p^2;t_{thr2},m_2^2) }
                       {(ds_2)/(dt)|_{t=t_{thr2}}} \,.
\end{eqnarray}

Some special mass cases have to be considered.
Double poles occur in $R(s)$ if $m_2^2=0$, in which case $s_3=s_4$, or
if $\lambda(m_2^2,m_5^2,m_3^2)=0$, in which case $s_1=s_2$.
In the case of $m_1^2=0$ the partial integration (\ref{DT3par}) is modified.
For $T^{(3b)}$, these conditions amount to $m_1^2=0$,
$\lambda(m_1^2,m_4^2,m_3^2)=0$ or $m_2^2=0$.
The decomposition into partial fractions
leads then to modified results involving functions
\begin{equation}
  \int\limits _{(\sqrt{t}+m)^2}^\infty \frac{ds}{(s-s_i-i\epsilon)^2}
             \frac{\sqrt{\lambda(s,t,m^2)}}{s} =
       \frac{\partial B_0(s_i;t,m^2)}{\partial s_i} \,.
\end{equation}
These derivatives of the $B_0$-function are finite in four dimensions.
Another modification and considerable simplification of the decomposition
into partial fractions occurs if $m_3^2=0$.

In our evaluation of the three-particle cut contributions the symmetry of
the master diagram with respect to an simultaneous interchange of the masses
$m_1^2 \leftrightarrow m_4^2$ and $m_2^2 \leftrightarrow m_5^2$ is
hidden.
Broadhurst has evaluated the master diagram for many cases
of physical interest \cite{Broadhurst}. Note that all these cases
can be understood to belong to the special cases specified above, if
one takes the symmetries of the master diagram into account.

\section{Numerical comparisons}
\label{Numerics}

The formulae (\ref{T2}) and (\ref{T12345-3})
provide an efficient algorithm for a numerical calculation of
the master diagram in the general
mass case.
The results have been checked by a comparison with Kreimer's
two-dimensional integral representation
\cite{Kreimer}. We implemented both algorithms with C++ and
performed a test for more than $150000$ different values of the parameters.
This should be sufficient to check the parameter region.

For the one-dimensional integral representation an
adaptive Gauss-Kronrod algorithm, implemented in the QUADPACK-routines
\cite{QUADPACK} proved suitable. For the two-dimensional integral we did not
succeed with a Gauss integration. But following ref.~\cite{Berends}, we
used the VEGAS-routine \cite{VEGAS} which implements
an adaptive Monte Carlo integration algorithm.

The following tables are intended to compare typical results of
the two algorithms. We have chosen a parameter region where $p^2$ has
a similar magnitude as the internal masses, where the integral
representation is most useful to handle.
Table \ref{tab-one-dim} gives the results of the one-dimensional integral
representation, table \ref{tab-two-dim} of the two-dimensional integral
representation.
In each row the results for the real and the imaginary part, the
reported error and the CPU-time are presented.
To get a better estimate of the
real errors and some impression about the convergence qualities of the
algorithms, two runs have been performed for each parameter set.
In case of the one-dimensional representation the required accuracy
has been $10^{-4}$ for the first run and $5 \times 10^{-6}$ for the second run.
In the case of the two-dimensional representations we used $10^5$
function evaluations for the first run and $5 \times 10^5$ function evaluations
for the second run.
The numerical results for the
imaginary parts are reported even in those regions, where
$p^2$ is below all thresholds, because the magnitude of the imaginary part
gives another estimate for the errors.

The precision of the one-dimensional integral representation can be increased
further by introducing asymptotic formulae for the integrand at large
values of the integration variable.
With one-dimensional integral representations
for other two-loop self-energy integrals \cite{BBBB1,BBBB2} we made the
experience that simple asymptotic formulae increase
the maximal precision by several
orders of magnitude. We are going to implement this also for the
algorithm presented in this paper. A first version of the program
can be requested from the authors.

The programs have been run
on a workstation DEC 3000 AXP. Note that the error reported by
the VEGAS-routine gives the size of a standard deviation,
while the error reported by
the QUADPACK-routine is not so clear defined, but is in general highly
over-estimated. This can be observed from the tables by a comparison
of the results with different precision.

\begin{table}[htb]
\begin{center}
\begin{tabular}{|r|r|r|r|r|r|} \hline
\multicolumn{1}{|c}{$p^2$}  & \multicolumn{1}{|c}{A} &
\multicolumn{1}{|c}{B} & \multicolumn{1}{|c}{C} &
\multicolumn{1}{|c}{D} & \multicolumn{1}{|c|}{E} \\
\hline
$0.1$&$<$&$-2.87238e-01$&$-1.84514e-09$&$4.9e-05$&$0.6$ \\
$0.1$&$<$&$-2.87238e-01$&$-1.57367e-10$&$2.5e-06$&$0.8$ \\
$0.5$&$<$&$-2.94592e-01$&$-1.89266e-09$&$5.0e-05$&$0.7$ \\
$0.5$&$<$&$-2.94592e-01$&$-1.61303e-10$&$2.5e-06$&$0.8$ \\
$1.0$&$<$&$-3.04522e-01$&$-1.95521e-09$&$5.0e-05$&$0.7$ \\
$1.0$&$<$&$-3.04521e-01$&$-1.66484e-10$&$2.5e-06$&$0.8$ \\
$5.0$&$<$&$-4.52521e-01$&$-2.62945e-09$&$5.6e-05$&$0.7$ \\
$5.0$&$<$&$-4.52520e-01$&$-2.22337e-10$&$2.8e-06$&$0.8$ \\
$10.0$&$>$&$-4.88154e-01$&$-3.53218e-01$&$4.9e-05$&$0.7$ \\
$10.0$&$>$&$-4.88153e-01$&$-3.53217e-01$&$2.3e-06$&$0.9$ \\
$50.0$&$>$&$1.73902e-01$&$-1.18080e-01$&$4.3e-05$&$1.4$ \\
$50.0$&$>$&$1.73901e-01$&$-1.18080e-01$&$2.2e-06$&$1.8$ \\
\hline
\end{tabular}
\end{center}
\caption[]{The master integral $\TM(p^2;m_1^2,m_2^2,m_3^2,m_4^2,m_5^2)$,
calculated with the one-dimensional integral representation.
The masses are $m_1^2=1$, $m_2^2=2$, $m_3^2=3$, $m_4^2=4$, $m_5^2=5$.
A: '$<$' indicates, that $p^2$ is below all thresholds, i.e.~the imaginary part
is zero.
B: Numerical result for the real part.
C: Numerical result for the imaginary part.
D: Error reported by the QUADPACK-routine.
E: CPU-time in seconds.
}
\label{tab-one-dim}
\end{table}

\begin{table}[htb]
\begin{center}
\begin{tabular}{|r|r|r|r|r|r|} \hline
\multicolumn{1}{|c}{$p^2$}  & \multicolumn{1}{|c}{A} &
\multicolumn{1}{|c}{B} & \multicolumn{1}{|c}{C} &
\multicolumn{1}{|c}{D} & \multicolumn{1}{|c|}{E} \\
\hline
$0.1$&$<$&$-2.88941e-01$&$4.46110e-03$&$3.4e-03$&$16$ \\
$0.1$&$<$&$-2.87005e-01$&$-2.17356e-04$&$7.0e-04$&$84$ \\
$0.5$&$<$&$-2.94846e-01$&$4.17190e-04$&$1.6e-03$&$16$ \\
$0.5$&$<$&$-2.94795e-01$&$-4.88930e-05$&$3.5e-04$&$85$ \\
$1.0$&$<$&$-3.05219e-01$&$8.17018e-07$&$1.3e-03$&$17$ \\
$1.0$&$<$&$-3.04930e-01$&$-3.78437e-04$&$2.8e-04$&$85$ \\
$5.0$&$<$&$-4.53058e-01$&$-8.90271e-04$&$8.3e-04$&$17$ \\
$5.0$&$<$&$-4.52409e-01$&$-9.65061e-05$&$1.8e-04$&$86$ \\
$10.0$&$>$&$-4.88270e-01$&$-3.52664e-01$&$4.8e-04$&$17$ \\
$10.0$&$>$&$-4.88250e-01$&$-3.53326e-01$&$1.1e-04$&$86$ \\
$50.0$&$>$&$1.73864e-01$&$-1.18080e-01$&$6.9e-05$&$17$ \\
$50.0$&$>$&$1.73907e-01$&$-1.18072e-01$&$1.5e-05$&$85$ \\
\hline
\end{tabular}
\end{center}
\caption[]{The master integral $\TM(p^2;m_1^2,m_2^2,m_3^2,m_4^2,m_5^2)$,
calculated with the two-dimensional integral representation.
The masses are $m_1^2=1$, $m_2^2=2$, $m_3^2=3$, $m_4^2=4$, $m_5^2=5$.
A: '$<$' indicates, that $p^2$ is below all thresholds, i.e.~the imaginary part
is zero.
B: Numerical result for the real part.
C: Numerical result for the imaginary part.
D: Error reported by the VEGAS-routine.
E: CPU-time in seconds.
}
\label{tab-two-dim}
\end{table}

As expected, the one-dimensional integral representation is
considerably faster. The difference is more important if higher
accuracy is required. Note that the CPU-time for the one-dimensional
integration is nearly the same for the two runs for all values of $p^2$,
while the accuracy has increased by a factor of $20$. For the
two-dimensional integration, on the other hand, a difference in the
computing-time
by a factor $5$ causes about the same increase in accuracy.
While the one-dimensional integration gives results accurate to about
$6$ digits (regarding the real error) in about $1$ second,
the two-dimensional integration
needs about $85$ seconds to achieve an accuracy of between $3$ and $4$
digits, depending on the parameters.

\section{Conclusion}

For scalar two-loop self-energy integrals analytic expressions have
been presented in the literature for many special cases
\cite{Broadhurst,ScharfTausk}. Asymptotic
expansions and Taylor series are another important approach to
these integrals, see e.g.~\cite{asymptotic}.
Furthermore, for diagrams containing a self-energy subloop,
series representations have been presented \cite{Buza}-\cite{BBBB2}
which are analytical results for the general mass case.
No comparable analytical result has so far been found for the master
diagram in the general mass case. Anyhow, series representations always
have a restricted region of convergence, i.e.~in each particular region
up to the closest
threshold or pseudo-threshold. In the general case of the master diagram
with its many parameters this will lead to considerable complications
concerning representations in terms of generalized hypergeometric functions
or in terms of Taylor series.

In this paper we presented analytical results for the imaginary part of the
master diagram $\TM$ and a one-dimensional integral representation
for $\TM$.
In all our results the two- and three-particle cut contributions are
separated, which gives some additional insight into the analytical structure
of the function $\TM$.
Together with corresponding results \cite{BBBB1,BBBB2}
for the other fundamental
two-loop self-energy integrals and with a tensor reduction formalism
for the two-loop self-energy integrals \cite{Weig},
efficient algorithms are now
available for the calculation of two-loop self-energies in the general
mass case.
The numerical implementations should be
sufficient to perform complete
calculations in the standard model at the two-loop level,
e.g.~to evaluate the fermion or gauge-boson self-energies.

\section*{Acknowledgements}
The authors are grateful to F.~A.~Berends, M.~Buza and G.~Weiglein for useful
discussions.


\begin{thebibliography}{99}
%
\bibitem{MZ}A.~Blondel, {\em talk given at the 1994 Zeuthen workshop on
elementary particle physics}, Teupitz (1994).
%
\bibitem{Buza}F.~A.~Berends, M.~Buza, M.~B\"{o}hm and R.~Scharf,
{\em Z.~Phys.}~C63 (1994) 227.
%
\bibitem{BBBB1}S.~Bauberger, F.~A.~Berends, M.~B\"{o}hm and M.~Buza,
{\em Leiden preprint INLO-PUB-9/94},
{\em W\"urzburg preprint UWITP-2/94},
to appear in {\em Nucl. Phys.}~B (hep-ph/9409388).
\bibitem{BBBB2}
S.~Bauberger, F.~A.~Berends, M.~B\"{o}hm, M.~Buza and G.~Weiglein,
{\em Nucl. Phys. B (Proc. Suppl.)}~37B (1994) 95 (hep-ph/9406404).
%
\bibitem{Broadhurst}D.~J.~Broadhurst, {\em Z.Phys.}~C47 (1990) 115.
%
\bibitem{ScharfTausk}R.~Scharf and J.~B.~Tausk,
{\em Nucl.~Phys.~}B~412 (1994) 423.
%
\bibitem{Ghinculov}
A.~Ghinculov, J.~J.~van der Bij, {\em Massive two--loop diagrams:
The Higgs propagator}, Freiburg preprint THEP 94/05 (hep-ph/9405418).
%
\bibitem{asymptotic}
A.~I.~Davydychev, J.~B.~Tausk, {\em Nucl. Phys.}~B397 (1993) 123. \\
D.~J.~Broadhurst, J.~Fleischer, O.~V.~Tarasov,
{\em Z. Phys.}~C60 (1993) 287 (hep-ph/9304303). \\
A.~I.~Davydychev, V.~A.~Smirnov, J.~B.~Tausk, {\em Nucl. Phys.}~B410
(1993) 325 (hep-ph/9307371). \\
J.~Fleischer, O.~V.~Tarasov, {\em Z. Phys.}~C64 (1994) 413
(hep-ph/9403230). \\
J.~Fleischer, O.~V.~Tarasov, {\em Application of conformal mapping and
Pad\'e approximants ($\omega$ P's) to the calculation of various two-loop
Feynman diagrams}, (hep-ph/9407235). \\
F.~A.~Berends, A.~I.~Davydychev, V.~A.~Smirnov, J.~B.~Tausk,
{\em Zero-threshold expansions of two-loop self-energy diagrams},
Leiden-preprint INLO-PUB-15/94, DTP-94/86 (hep-ph/9410232).
%
\bibitem{Scharf}R.~Scharf, Diploma Thesis, W\"urzburg (1991).
%
\bibitem{Kreimer}D.~Kreimer,
{\em Phys.~Lett.}~B273 (1991) 277.
%
\bibitem{Berends}F.~A.~Berends and J.~B.~Tausk,
{\em Nucl.~Phys.~}~B421 (1994) 456.
%
\bibitem{Cutkosky}R.~E.~Cutkosky, {\em J.~Math.~Phys.}~1 (1960) 429.\\
G.'t~Hooft, M.~Veltman, Diagrammar, CERN Yellow Report 73-9.
%
\bibitem{Itzykson}C.~Itzykson, J.~B.~Zuber, {\em Quantum field theory},
McGraw-Hill, New York (1980).
%
\bibitem{tHooft2}G.~'t~Hooft, M.~Veltman, {\em Nucl. Phys.}~B153 (1979) 365.
%
\bibitem{anomalous}S.~Mandelstam, {\em Phys. Rev. Letters}~4 (1960) 84. \\
A.~O.~Barut, {\em The theory of the scattering matrix},
The Macmillan Company, New York (1967).
%
\bibitem{Erd}A.~Erd\'{e}ly, W.~Magnus, F.~Oberhettinger, F.~G.~Tricomi,
{\em Higher Transcendental Functions}, vol. II, McGraw-Hill,
New York (1953).
%
\bibitem{hypergeometric}H.~Exton,
{\em Multiple Hypergeometric Functions and Applications},
Ellis Horwood Limited, Chicester (1976). \\
P.~Appell, J.~Kamp\'{e} de F\'{e}riet,
{\em Fonctions Hyperg\'{e}om\'{e}triques}, Gauthier-Villars, Paris (1926). \\
H.~M.~Srivastava, P.~W.~Karlson,
{\em Multiple Gaussian Hypergeometric Series}, Halstead Press, John Wiley,
New York (1985).
%
\bibitem{QUADPACK}R.~Piessens, E.~de~Doncker-Kapenga, C.~W.~\"Uberhuber,
D.~K.~Kahanger, {\em QUADPACK, A Subroutine Package for Automatic Integration},
Springer, Berlin (1983).
%
\bibitem{VEGAS}G.~P.~Lepage, {\em J.~Comp.~Phys.}~27 (1978) 192;
Cornell preprint CLNS-80/447 (1980).
%
\bibitem{Weig}G.~Weiglein, R.~Scharf, M.~B\"ohm,
{\em Nucl.~Phys.}~B416 (1994) 606.
%
\end{thebibliography}
\end{document}